\documentstyle[12pt]{article}

\font\titlefont=cmbx10 scaled \magstep4
\begin{document}

\begin{flushright}
\vspace*{-2cm}
gr-qc/9410047  \\ TUTP-94-15  \\  October 1994
\vspace*{1.5cm}
\end{flushright}

\begin{center}
{\titlefont GRAVITONS AND LIGHTCONE\\
\vskip 0.2in
FLUCTUATIONS }
\vskip .5in
L.H. Ford \\
\vskip .3in
Institute of Cosmology\\
Department of Physics and Astronomy\\
Tufts University\\
Medford, Massachusetts 02155\\
\end{center}

\vskip 0.5in

\begin{abstract}
Gravitons in a squeezed vacuum state, the natural result of quantum creation
in the early universe or by black holes, will introduce metric fluctuations.
These metric fluctuations will introduce fluctuations of the lightcone. It
is shown that when the various two-point functions of a quantized field are
averaged over the metric fluctuations, the lightcone singularity disappears
for distinct points. The metric averaged functions remain singular in the limit
of coincident points. The metric averaged retarded Green's function for a
massless field becomes a Gaussian which is nonzero both inside and outside
of the classical lightcone. This implies some photons propagate faster than
the classical light speed, whereas others propagate slower. The possible
effects of metric fluctuations upon one-loop quantum processes are discussed
and illustrated by the calculation of the one-loop electron self-energy.
\end{abstract}
\newpage
\baselineskip=14pt

\section{Introduction}

      It was conjectured several years ago by Pauli\cite{Pauli} that
the ultraviolet divergences of quantum field theory might be removed
in a theory in which gravity is quantized. The basis of Pauli's conjecture
was the observation that these divergences arise from the lightcone
singularities of two-point functions, and that quantum fluctuations of
the spacetime metric ought to smear out the lightcone, possibly removing
these singularities. This conjecture was discussed further by
Deser\cite{Deser},
in the context of a path integral approach to the quantization of gravity,
and by Isham, Salam, and Strathdee\cite{ISS}.
However, there seems to have been little progress on this question in the
intervening years. Indeed, it is well known that perturbative quantum
gravity, far from being a universal regulator, is afflicted with
nonrenormalizable infinities of its own.
     In the present work, the issue of lightcone fluctuations will be
examined in a context where they are produced by gravitons propagating on
a flat background. We assume that the gravitons are in a squeezed vacuum
state, which is the appropriate state for relic gravitons created by quantum
particle creation processes in the early universe\cite{Grishchuk} or by black
hole evaporation. More generally, a squeezed vacuum state is the quantum state
which arises in any quantum particle creation process in which the state
of the created particles is an in-vacuum state represented in an out-Fock
space. It will be
shown that averaging over the metric fluctuations associated with such
gravitons has the effect of smearing out the lightcone.

      It should be noted that the metric fluctuations being considered in
this paper are distinct from those due to fluctuations in the energy-momentum
tensor of the source\cite{F82,Kuo}. It is possible for the energy density,
for example, to exhibit large fluctuations. This arises in the Casimir effect
and in quantum states in which the expectation value of the energy density
is negative. This means that the gravitational field of such a system is not
described by a fixed classical metric, but rather by a fluctuating metric.
However, these metric fluctuations are ``passive'' in the sense that they
are driven by fluctuations in the degrees of freedom of the matter field.
In contrast, the metric fluctuations due to gravitons in a squeezed state
are ``active'' fluctuations produced by quantized degrees of freedom of
the gravitational field itself.

      In Section ~\ref{sec:aveGF}, the retarded, Hadamard, and Feynman
functions will be averaged over metric fluctuations. The resulting smearing
of the lightcone is also discussed. The average of the square of the
Feynman propagator for a scalar field is also calculated.
The results are given in terms of the mean
square of the squared geodesic separation between points. In Section
{}~\ref{sec:form}, this quantity is calculated explicitly for various cases.
In this section, gravitons in an expanding universe are also discussed,
and some estimates for the present background of relict gravitons are
given. The one-loop electron self-energy is calculated in the presence
of metric fluctuations is calculated and discussed in Section
{}~\ref{sec:oneloop}. The results of the paper are summarized and discussed
in Section ~\ref{sec:summary}.

\section{Averaging Two-Point Functions over Metric Fluctuations}
\label{sec:aveGF}
\subsection{The Retarded Green's Function}

      Let us consider a flat background spacetime with a linearized
perturbation $h_{\mu\nu}$ propagating upon it. Thus the spacetime metric
may be written as
\begin{equation}
ds^2 = g_{\mu\nu}dx^\mu dx^\nu = (\eta_{\mu\nu} +h_{\mu\nu})dx^\mu dx^\nu
= dt^2 -d{\bf x}^2 + h_{\mu\nu}dx^\mu dx^\nu \, .  \label{eq:metric}
\end{equation}
In the unperturbed spacetime, the square of the geodesic separation of
points $x$ and $x'$ is $2\sigma_0 =(x-x')^2 = (t-t')^2 -({\bf x}-{\bf x}')^2$.
In the presence of the
perturbation, let this squared separation be $2\sigma$, and write
\begin{equation}
\sigma= \sigma_0 + \sigma_1 + O(h_{\mu\nu}^2),
\end{equation}
so $\sigma_1$ is the  shift in $\sigma$ to first order in $h_{\mu\nu}$.

Let us consider the retarded Green's function for a massless scalar field.
In flat spacetime, this function is
\begin{equation}
G_{ret}^{(0)}(x-x') = {{\theta(t-t')}\over {4\pi}} \delta(\sigma_0)\, ,
\end{equation}
which has a delta-function singularity on the future lightcone and is zero
elsewhere. In the presence of a classical
metric perturbation, the retarded Green's function has its delta-function
singularity on the perturbed lightcone, where $\sigma=0$. In general,
it may also become nonzero on the interior of the lightcone due to
backscattering off of the curvature. However, we are primarily interested
in the behavior near the new lightcone, and so let us replace
$G_{ret}^{(0)}(x-x')$ by
\begin{equation}
G_{ret}(x,x') = {{\theta(t-t')}\over {4\pi}} \delta(\sigma)\, .
                            \label{eq:gret0}
\end{equation}
We are assuming that the curved space Green's functions have the Hadamard form,
in which case their leading asymptotic behavior near the lightcone is the same
as in flat space \cite{Fulling}.
One may regard this assumption as a restriction
on the physically allowable quantum states.
If we terminate the expansion of $\sigma$ at first order (higher orders will
be discussed below), then Eq. (\ref{eq:gret0}) may be expressed as
\begin{equation}
G_{ret}(x,x') = {{\theta(t-t')}\over {8\pi^2}} \int_{-\infty}^{\infty}
                d\alpha\, e^{i\alpha \sigma_0}\, e^{i\alpha \sigma_1}\, .
                                                         \label{eq:gretrep}
\end{equation}

We now replace the classical metric perturbations by gravitons in a squeezed
vacuum state $|\psi\rangle$. Then $\sigma_1$ becomes a quantum operator
which is linear
in the graviton field operator, $h_{\mu\nu}$. A squeezed vacuum state
is a state such that $\sigma_1$ may be decomposed into positive and negative
frequency parts. Thus we may find $\sigma_1^{+}$ and $\sigma_1^{-}$ so that
\begin{equation}
\sigma_1^{+} |\psi\rangle =0, \qquad \langle \psi| \sigma_1^{-}=0\, ,
                                            \label{eq:posfreq}
\end{equation}
where $\sigma_1 = \sigma_1^{+} + \sigma_1^{-}$. In terms of annihilation
and creation operators, $\sigma_1^{+} = \sum_j a_j f_j$ and $\sigma_1^{-} =
\sum_j a^\dagger_j f^*_j$, where the $f_j$ are mode functions. We now write
\begin{equation}
e^{i\alpha\sigma_1} = e^{i\alpha(\sigma_1^{+} + \sigma_1^{-})} =
e^{i\alpha\sigma_1^{-}}
e^{-{1\over 2}\alpha^2 [\sigma_1^{+}, \sigma_1^{-}]} e^{i\alpha\sigma_1^{+}}\,.
                                         \label{eq:expop}
\end{equation}
In the second step we used the Campbell-Baker-Hausdorff formula, that
$e^{A+B} = e^A e^{\frac{1}{2}[A,B]} e^B$ for any pair of operators $A$ and $B$
that each commute with their commutator, $[A,B]$.
We now take the expectation value of this expression and use the
facts that $e^{i\alpha\sigma_1^{+}}|\psi\rangle = |\psi\rangle$ and
$\langle \psi|e^{i\alpha\sigma_1^{-}}
= \langle \psi|$, which follow immediately from Eq. (~\ref{eq:posfreq})
if the exponentials are expanded in
a power series. Finally, we use $[\sigma_1^{+}, \sigma_1^{-}] =\sum_j f_j f^*_j
= \langle {\sigma_1}^2 \rangle$ to write
\begin{equation}
\Bigl\langle e^{i\alpha \sigma_1} \Bigr\rangle =
e^{-{1\over 2}\alpha^2 \langle \sigma_1^2 \rangle} \, . \label{eq:expav}
\end{equation}
Thus when we average over the metric fluctuations, the retarded Green's
function is replaced by its quantum expectation value:
\begin{equation}
\Bigl\langle G_{ret}(x,x') \Bigr\rangle =
{{\theta(t-t')}\over {8\pi^2}} \int_{-\infty}^{\infty}
                          d\alpha \,e^{i\alpha \sigma_0} \,
e^{-{1\over 2}\alpha^2 \langle \sigma_1^2 \rangle} \, .
\end{equation}
The expectation value of $\sigma_1^2$ is formally divergent. However, in flat
spacetime this divergence may be removed by subtraction of the expectation
value in the Minkowski vacuum state. Henceforth, we will take
$\langle \sigma_1^2 \rangle$ to denote this renormalized expectation value.

The above integral converges only if $\langle \sigma_1^2 \rangle > 0$, in which
case it may be evaluated to yield
\begin{equation}
\Bigl\langle G_{ret}(x,x') \Bigr\rangle =
{{\theta(t-t')}\over {8\pi^2}} \sqrt{\pi \over {2\langle \sigma_1^2 \rangle}}
\; \exp\Bigl(-{{\sigma_0^2}\over {2\langle \sigma_1^2 \rangle}}\Bigr)\, .
                                           \label{eq:retav}
\end{equation}
Note that this averaged Green's function is indeed finite at $\sigma_0 =0$
provided that $\langle \sigma_1^2 \rangle \not= 0$. Thus the lightcone
singularity has been smeared out.
Note that the smearing occurs in both the timelike and spacelike directions.

This smearing may be interpreted as due to the fact that photons may be
either slowed down or boosted by the metric fluctuations. Photon propagation
now becomes a statistical phenomenon; some photons travel slower than
light on the classical spacetime, whereas others travel faster. We have now
the possibility of ``faster than light'' signals. This need not cause any
causal paradoxes, however, because the system is no longer Lorentz invariant.
The graviton state defines a preferred frame of reference. The usual argument
linking superluminal signals with causality violation assumes Lorentz
invariance \cite{Pirani}.

     The effects of lightcone fluctuations upon photon propagation are
in principle observable. Consider a source which emits evenly spaced pulses.
An observer at a distance $D$ from the source will detect pulses whose spacing
varies by an amount of the order of $\Delta t$. For a pulse which is delayed by
time $\Delta t$,
\begin{equation}
\sigma = {1\over 2}[(D +\Delta t)^2 -D^2] \approx D \Delta t \, ,
   \qquad \Delta t \ll D \, .
\end{equation}
Thus the typical time delay or advance is of the order of
\begin{equation}
\Delta t \approx  {{\sqrt{\langle \sigma_1^2 \rangle}}\over D} \, .
                                         \label{eq:delt}
\end{equation}
This effect leads to the broadening of spectral lines. The observer will
detect a line which is broadened in wavelength by $\Delta \lambda =\Delta t$.
Some observational aspects of this effect will be discussed in more detail
in Section \ref{sec:form}.

     Note that it is essential that the gravitons be in a nonclassical
state, such as a squeezed vacuum, in order to obtain lightcone smearing.
Gravitons in a coherent state will represent a classical gravity wave.
In this case, the retarded Green's function will still have a delta function
singularity in the lightcone of the perturbed spacetime.

   In the above calculation of $\bigl\langle G_{ret}(x,x') \bigr\rangle$,
the expansion of $\sigma$ was truncated after the first order. However,
it is of interest to consider the effect of second order terms. This is
particularly pertinent in view of the fact that the crucial corrections
involve $\langle \sigma_1^2 \rangle$, which is itself second order in
$h_{\mu\nu}$ \cite{Boulware}. We now write
\begin{equation}
\sigma= \sigma_0 + \sigma_1 + \sigma_2 + O(h_{\mu\nu}^3),
\end{equation}
so that $\sigma_2$ is the second order correction. We now wish to include this
correction in the calculation of $\bigl\langle G_{ret}(x,x') \bigr\rangle$.
Let us first write
\begin{equation}
\sigma_2 = :\sigma_2: + \langle\sigma_2\rangle \, ,
\end{equation}
where the colons denote normal ordering with respect to the state
$|\psi\rangle$, and the expectation value is understood to be in this state.
Equation (\ref{eq:expop}) is now replaced by
\begin{equation}
e^{i\alpha(\sigma_1 +\sigma_2)} =
e^{i\alpha\sigma_1^{-}}
e^{-{1\over 2}\alpha^2 [\sigma_1^{+}, \sigma_1^{-}]} e^{i\alpha\sigma_1^{+}}
e^{i\alpha\langle\sigma_2\rangle}\,e^{i\alpha:\sigma_2:} \,.
                                         \label{eq:expop2}
\end{equation}
Here we have ignored all terms which are of third order or higher, including
those which arise when $:\sigma_2:$ is commuted past $\sigma_1^{\pm}$. We
use the fact that
\begin{equation}
e^{i\alpha:\sigma_2:} |\psi \rangle = |\psi\rangle \,,
\end{equation}
to write the analog of Eq.~(\ref{eq:expav}):
\begin{equation}
\Bigl\langle e^{i\alpha (\sigma_1 +\sigma_2)} \Bigr\rangle =
e^{i\alpha\langle\sigma_2\rangle
     -{1\over 2}\alpha^2 \langle \sigma_1^2 \rangle} \, . \label{eq:expav2}
\end{equation}
As in the case of $\langle \sigma_1^2 \rangle$, we assume that
$\langle\sigma_2\rangle$ is a renormalized expectation value. Now the
metric averaged Green's function
becomes
\begin{equation}
\Bigl\langle G_{ret}(x,x') \Bigr\rangle =
{{\theta(t-t')}\over {8\pi^2}} \sqrt{\pi \over {2\langle \sigma_1^2 \rangle}}
\; \exp\Bigl(-{{\sigma_0^2 +\langle\sigma_2\rangle }\over
                                  {2\langle \sigma_1^2 \rangle}}\Bigr)\, .
                                           \label{eq:retav2}
\end{equation}
Comparison with Eq.~(\ref{eq:retav}) reveals that the effect of retaining the
$\sigma_2$ term is simply to shift slightly the position of the peak of the
Gaussian. Thus  $\langle\sigma_2\rangle$ enters in a different way from
$\langle \sigma_1^2 \rangle$, due to the different powers of $\alpha$ in
Eq.~(\ref{eq:expav2}). The same phenomenon would occur for the other functions
to be discussed below, so henceforth the $\sigma_2$ terms will be ignored.

    It should be noted that although we are expanding $\sigma$ in powers
of the metric perturbation $h_{\mu\nu}$, the averaging procedure used to
obtain $\bigl\langle G_{ret}(x,x') \bigr\rangle$ retains terms of all
orders in $h_{\mu\nu}$. This is essential in order to obtain nontrivial
results. We can think of this as an expansion of the {\it argument} of the
exponential functions in Eqs.~(\ref{eq:gretrep}) or (\ref{eq:expop2}) but not
of the functions themselves. This seems to be self-consistent in that
retaining successively higher terms in $\sigma$ leads to small changes
in the form of the results, as we saw in going from Eq.~(\ref{eq:retav}) to
Eq.~(\ref{eq:retav2}).

\subsection{The Hadamard Function}

      In addition to the retarded and advanced Green's functions discussed
in the previous subsection, there are several other singular functions
in quantum field theory which can be expressed as vacuum expectation values of
products of field operators. In particular, the {\it Hadamard function} for
a scalar field $\phi$ is defined as
\begin{equation}
G_1 (x,x') \equiv \langle 0|\phi(x) \phi(x')+ \phi(x') \phi(x)|0\rangle,
\end{equation}
where $|0\rangle$ is the vacuum state. In the massless case in flat
spacetime, it has the explicit form:
\begin{equation}
G_1(x,x') = -{1 \over {4\pi^2 \sigma}}. \label{eq:Had}
\end{equation}
Recall that $\sigma$ is one-half of the square of the geodesic distance
between $x$ and $x'$, and  in flat spacetime, $\sigma = {1\over 2} (x-x')^2$.
Even in the massive case, and/or in curved spacetime, Eq. (\ref{eq:Had})
gives the asymptotic behavior of $G_1(x,x')$ near the lightcone.
As in the case of the retarded Green's function, we now wish to replace
$\sigma$ by $\sigma_0 +\sigma_1$ and take the quantum expectation value of
the result. Let us use the identities
\begin{equation}
\int_0^\infty d\alpha \, e^{i\alpha x} = {i \over x} +\pi \delta(x),
                                          \label{eq:delta}
\end{equation}
and
\begin{equation}
\int_0^\infty d\alpha \, e^{-i\alpha x} = -{i \over x} +\pi \delta(x),
\end{equation}
to write
\begin{equation}
{1 \over {\sigma_0 +\sigma_1}} =
     -{i \over 2} \int_0^\infty d\alpha \,
\bigl[e^{i(\sigma_0 +\sigma_1)\alpha} -e^{-i(\sigma_0 +\sigma_1)\alpha}\bigr].
\end{equation}
Now use Eq. (\ref{eq:expav}) to take the expectation value of the above
expression and write
\begin{equation}
\Bigl\langle G_1 (x,x') \Bigr\rangle =
 -{1 \over {4\pi^2}} \Biggl\langle{1 \over {(\sigma_0 +\sigma_1)}}\Biggr\rangle
=-{1 \over {4\pi^2}} \int_0^\infty d\alpha \, \sin \sigma_0\alpha \,\,
      e^{-{1\over 2}\alpha^2 \langle \sigma_1^2 \rangle}. \label{eq:Hadav1}
\end{equation}
This expression gives us the Hadamard function averaged over metric
fluctuations
for the case that $\langle \sigma_1^2 \rangle > 0$.

   Let us examine the asymptotic forms of this result. Near the classical
lightcone, $\sigma_0 \rightarrow 0$. If we expand the integrand of the
above expression to lowest order in $\sigma_0$, and perform the integration,
we find that
\begin{equation}
\Bigl\langle G_1 (x,x') \Bigr\rangle \sim
-{{\sigma_0}\over {4\pi^2 \langle \sigma_1^2 \rangle}},
                                 \qquad \sigma_0 \rightarrow 0.
\end{equation}
Thus the lightcone singularity is removed so long as
$\langle \sigma_1^2 \rangle \not= 0$, which will generally be the case for
non-coincident points. Equation (\ref{eq:Hadav1}) may be rewritten as
\begin{equation}
\Bigl\langle G_1 (x,x') \Bigr\rangle =
 -{1 \over {4\pi^2 \sigma_0}}
\Bigl[ 1 -{{\langle \sigma_1^2 \rangle}\over {\sigma_0^2}}
\int_0^\infty dt\, t \, \cos t \,\,
\exp\Bigl(-{{\langle \sigma_1^2 \rangle t^2}\over {2 \sigma_0^2}}\Bigr)\Bigr].
\end{equation}
In the limit that $\sigma_0^2 \gg \langle \sigma_1^2 \rangle$, the second term
above is negligible and we recover the classical form of $G_{1}$ :
\begin{equation}
\Bigl\langle G_{1}(x,x') \Bigr\rangle \sim -{1 \over{4\pi^2 \sigma_0}}.
                                              \label{eq:classlim}
\end{equation}

The above expression for $\Bigl\langle G_1 (x,x') \Bigr\rangle$ is valid
for $\langle \sigma_1^2 \rangle > 0$. We can, however, obtain an alternative
form valid for the case that $\langle \sigma_1^2 \rangle < 0$. To do so,
we use the representation
\begin{equation}
{1 \over {\sigma_0 +\sigma_1}} =
     \int_0^\infty d\alpha \, e^{-(\sigma_0 +\sigma_1)\alpha}.
\end{equation}
Now we have
\begin{equation}
\Bigl\langle G_1 (x,x') \Bigr\rangle =
 -{1 \over {4\pi^2}} \Biggl\langle{1 \over {(\sigma_0 +\sigma_1)}}\Biggr\rangle
=-{1 \over {4\pi^2}} \int_0^\infty d\alpha \, e^{-\sigma_0\alpha}\,
      e^{{1\over 2}\alpha^2 \langle \sigma_1^2 \rangle}. \label{eq:Hadav2}
\end{equation}
Near the lightcone, this quantity is finite:
\begin{equation}
\Bigl\langle G_1 (x,x') \Bigr\rangle \rightarrow
-{1 \over {4\pi^2}} \sqrt{\pi \over {2|\langle \sigma_1^2 \rangle|}},
                                 \qquad \sigma_0 \rightarrow 0.
\end{equation}
We may rewrite Eq. (\ref{eq:Hadav2}) as
\begin{equation}
\Bigl\langle G_1 (x,x') \Bigr\rangle =
 -{1 \over {4\pi^2 \sigma_0}}
\Bigl[ 1 +{{\langle \sigma_1^2 \rangle}\over {\sigma_0^2}}
\int_0^\infty dt\, t \, e^{-t} \,
\exp\Bigl({{\langle \sigma_1^2 \rangle t^2}\over {2 \sigma_0^2}}\Bigr)\Bigr].
                                                \label{eq:Hadav2b}
\end{equation}
{}From this form, we again obtain Eq. (\ref{eq:classlim}) when
$\sigma_0^2 \gg |\langle \sigma_1^2 \rangle|$.
Alternatively, Eq. (\ref{eq:Hadav2b}) may be derived by expanding
$(\sigma_0 +\sigma_1)^{-1}$ in a power
series in $\sigma_1$, using Wick's theorem to replace $\langle \sigma_1^{2n}
\rangle$ by $(2n-1)!! \,\langle \sigma_1^{2} \rangle^n$, and finally
resuming the result by Borel summation.

\subsection{The Feynman Propagator}

    The average of the Feynman propagator, $G_F$, over the metric fluctuations
can readily be obtained by combining the results of the previous two
subsections. We use the identity
\begin{equation}
G_F(x,x') = -{1\over 2}\bigl[G_{ret}(x,x') + G_{adv}(x,x')\bigr]
                        - {i\over 2}G_1 (x,x'),
                                              \label{eq:GF}
\end{equation}
and the fact that the advanced Green's function is related to the retarded
Green's function by
\begin{equation}
G_{adv}(x,x') = G_{ret}(x',x).
\end{equation}
We restrict our attention to the case that $\langle \sigma_1^2 \rangle > 0$,
both because it is only here that we have a formula for
$\Bigl\langle G_{ret}\Bigr\rangle$, and it is the case of greater physical
interest. Combining Eqs. (\ref{eq:retav}) and
(\ref{eq:Hadav1}), we obtain
\begin{equation}
\Bigl\langle G_{F}(x,x') \Bigr\rangle =
- { 1 \over {16\pi^2}} \sqrt{\pi \over {2\langle \sigma_1^2 \rangle}}
\; \exp\Bigl(-{{\sigma_0^2}\over {2\langle \sigma_1^2 \rangle}}\Bigr)
+{i \over {8\pi^2}} \int_0^\infty d\alpha \, \sin \sigma_0\alpha \,\,
   e^{-{1\over 2}\alpha^2 \langle \sigma_1^2 \rangle}\, .
                             \label{eq:Feyav}
\end{equation}
Again, this quantity is finite except in the coincidence limit,
$x' \rightarrow x$.

   Alternately, we can write
\begin{equation}
G_F(x,x') = {1\over {8\pi^2}}\biggl[ {i \over \sigma} -
 \pi \delta(\sigma) \biggr] =
  -{1\over {8\pi^2}}\int_0^\infty d\alpha \, e^{-i\alpha \sigma}\, .
                                            \label{eq:GFrep}
\end{equation}
Averaging this integral form for $G_F$ over metric fluctuations
yields
\begin{equation}
\Bigl\langle G_{F}(x,x') \Bigr\rangle =
-{1 \over {8\pi^2}} \int_0^\infty d\alpha \, e^{-i\sigma_0\alpha} \,\,
   e^{-{1\over 2}\alpha^2 \langle \sigma_1^2 \rangle}\, .
                             \label{eq:Feyav2}
\end{equation}
This form is equivalent to Eq.~(\ref{eq:Feyav}). Note that whereas the
real part of the above integral may be expressed in terms of elementary
functions, the imaginary part may not.

\subsection{The Square of the Feynman Propagator}
\label{sec:GF2}

    Earlier in this section, we obtained expressions for the various
singular functions averaged over metric fluctuations. However, the
Feynman diagrams for one-loop processes often involve products of at
least two Feynman propagators. Thus, if we wish to study the effect of
metric fluctuations upon these processes, we need an expression for quantities
such as $\Bigl\langle G_{F}^2 \Bigr\rangle$, the average of the square of the
Feynman propagator. We will again assume that
$\langle \sigma_1^2 \rangle >0$. We may use Eq.~(\ref{eq:GFrep}) to write
\begin{equation}
G_{F}^2 = {1 \over {(8\pi^2)}^2}\int_0^\infty d\alpha \, d\beta \,
                  e^{i(\alpha +\beta)\sigma}.
\end{equation}
If we set $\sigma= \sigma_0 + \sigma_1$, and average over the metric
fluctuations, the result is
\begin{equation}
\Bigl\langle G_{F}^2 \Bigr\rangle = {1 \over {(8\pi^2)}^2}
\int_0^\infty d\alpha \, d\beta \, e^{-i(\alpha +\beta)\sigma_0}
e^{-{1\over 2}(\alpha+\beta)^2 \langle \sigma_1^2 \rangle} \, .
\end{equation}
We next change the integration variables, first to polar coordinates
defined by $\alpha = \rho \cos \theta$ and $\beta = \rho \sin \theta$,
and then to a rescaled radial coordinate defined by $t =(\cos \theta +
\sin \theta)\rho$ :
\begin{eqnarray}
\Bigl\langle G_{F}^2 \Bigr\rangle &=& {1 \over {(8\pi^2)}^2}
 \int_0^{\pi \over 2} d\theta \int_0^\infty d\rho\, \rho\, e^{-i(\cos \theta +
\sin \theta)\sigma_0 \rho}
\exp[{-{1\over 2}(\cos \theta +
\sin \theta)^2 \langle \sigma_1^2 \rangle\rho^2}]         \nonumber \\
&=& {1 \over {(8\pi^2)}^2} \int_0^{\pi \over 2}
                          {{d\theta}\over (\cos \theta + \sin \theta)^2}
 \int_0^\infty dt\, t\, e^{-i\sigma_0 t}
e^{-{1\over 2}\langle \sigma_1^2 \rangle t^2} \, .
\end{eqnarray}
We now use the identity
\begin{equation}
\int_0^{\pi \over 2} {{d\theta}\over (\cos \theta + \sin \theta)^2} =1\, ,
\end{equation}
to write our result (for later use, it is convenient to relabel the integration
variable):
\begin{equation}
\Bigl\langle G_{F}^2 \Bigr\rangle = {1 \over {(8\pi^2)}^2}
 \int_0^\infty d\alpha\, \alpha\, e^{-i\sigma_0 \alpha}
e^{-{1\over 2}\langle \sigma_1^2 \rangle \alpha^2} \, . \label{eq:GFsqav}
\end{equation}
In this case, the imaginary part of the integral can be expressed in terms
of elementary functions to write
\begin{equation}
\Bigl\langle G_{F}^2 \Bigr\rangle =
{1 \over {64\pi^4}}\int_0^\infty d\alpha \,\alpha\, \cos\,\sigma_0\alpha \,\,
      e^{-{1\over 2}\alpha^2 \langle \sigma_1^2 \rangle}
+i{{\sqrt{2\pi}\sigma_0}\over {128\pi^4\langle \sigma_1^2 \rangle^{3\over
2}}}\:
\exp\Bigl(-{{\sigma_0^2}\over {2\langle \sigma_1^2 \rangle}}\Bigr)\, .
                                           \label{eq:GFsqav2}
\end{equation}
Again, these forms hold for the case that $\langle \sigma_1^2 \rangle >0$.
As in the case of the averaged Green's functions, this quantity is finite
on the classical lightcone, $\sigma_0 =0$, so long as the points are not
actually coincident, so $\langle \sigma_1^2 \rangle \not= 0$.

\section{Gravitons and the Form of $\langle \sigma_1^2 \rangle$ }
\label{sec:form}
\subsection{Gravitons in Flat Spacetime}

      So far, we have only assumed that the quantum state $|\psi\rangle$
of the gravitons is such that $\sigma_1$ can be decomposed into positive
and negative parts which satisfy Eq. (\ref{eq:posfreq}). However, we must
have more information about the state before we can determine the explicit
form of $\langle \sigma_1^2 \rangle$. Even the calculation of $\sigma_1$
for a given classical metric perturbation can be a difficult task, involving
the integration of the square root of Eq. (\ref{eq:metric}) along a
geodesic. However, as we are interested in gravitational wave perturbations,
we may simplify the analysis by the adoption of the transverse-tracefree
gauge, which is specified by the conditions
\begin{equation}
h^j_j = \partial_j h^{ij} = h^{0\nu} = 0\, . \label{eq:TT}
\end{equation}
In particular, $h_{\mu\nu}$ has purely spatial components, $h_{ij}$,
in a chosen coordinate system.
    Thus, in this gauge, a null geodesic is specified by
\begin{equation}
dt^2 = d{\bf x}^2 - h_{ij}dx^i dx^j \, ,
\end{equation}
and along a future-directed null geodesic, one has
\begin{equation}
dt = \sqrt{1 - h_{ij}n^i n^j }\, dr \approx
\left(1 - {1\over 2} h_{ij} n^i n^j \right)\,dr
 \, .
\end{equation}
Here $dr = |d{\bf x}|$, and $n^i ={{dx^i}/{dr}}$ is the
unit three-vector defining the spatial direction of the geodesic.
Thus the time interval $\Delta t$ and spatial interval $\Delta r = r_1 -r_0$
traversed by a null ray are related by
\begin{equation}
\Delta t = \Delta r -
{1\over 2}\int_{r_0}^{r_1} h_{ij} n^i n^j \,dr\,.
\end{equation}
Denote the right-hand side of the above expression by $\Delta \ell$, the
proper spatial distance interval between the endpoints. Now consider an
arbitrary pair of points (not necessarily null separated). The square of the
geodesic separation between these points is
\begin{equation}
2\sigma = (\Delta t)^2 - (\Delta \ell)^2 \approx (\Delta t)^2 - (\Delta r)^2
 + \Delta r \int_{r_0}^{r_1} h_{ij} n^i n^j \,dr\,,
\end{equation}
so
\begin{equation}
\sigma_1 = {1\over 2}\Delta r \int_{r_0}^{r_1}
                         h_{ij} n^i n^j\,dr\,.
\end{equation}
If we now treat $h_{ij}$ as a quantized metric perturbation, we obtain
a formula for $\langle \sigma_1^2 \rangle$ :
\begin{equation}
\langle \sigma_1^2 \rangle = {1\over 4}(\Delta r)^2
\int_{r_0}^{r_1} dr \int_{r_0}^{r_1} dr'
\:\, n^i n^j n^k n^m
\:\, \langle h_{ij}(x) h_{km}(x') \rangle \,.
\end{equation}
Here the graviton two-point function, $\langle h_{ij}(x) h_{km}(x') \rangle$,
is understood to be renormalized, so that it is finite when $x=x'$ and vanishes
when the quantum state of the gravitons is the vacuum state.

    Of particular interest is the case where only modes with wavelengths long
compared to $\Delta r$ are excited, so the two-point function is approximately
constant in both variables. Then
\begin{equation}
\langle \sigma_1^2 \rangle \approx {1\over 4}\langle h_{ij} h_{km}\, \rangle
\Delta x^i\Delta x^j \Delta x^k\Delta x^m \, ,
\end{equation}
where $\Delta x^i = ({{dx^i}/{dr}})\, \Delta r$ is the spatial coordinate
separation of the endpoints.
 In this frame of reference,
$\langle \sigma_1^2 \rangle$ will depend only upon $\Delta x^i$.

      We may illustrate the calculation of $\langle \sigma_1^2 \rangle$
more explicitly. The field operator $h_{\mu\nu}$ may be expanded in terms
of plane waves as
\begin{equation}
h_{\mu\nu} = \sum_{{\bf k},\lambda}\, [a_{{\bf k}, \lambda}
 e_{\mu\nu} ({{\bf k}, \lambda}) f_{\bf k} + H.c. ],
\end{equation}
where H.c. denotes the Hermitian conjugate, $\lambda$ labels the polarization
states,
$f_{\bf k} = (2\omega V)^{-{1\over 2}}  e^{i({\bf k \cdot x} -\omega t)}$
is a box normalized mode function, and the $e_{\mu\nu} ({{\bf k}, \lambda})$
are polarization tensors. (Here units in which $32\pi G =1$, where $G$ is
Newton's constant are used.) Let us consider the particular case of gravitons
in a squeezed vacuum state of a single linearly polarized plane wave mode.
Let the mode have
frequency $\omega$ and be propagating in the $+z$ direction. Take the
polarization tensor to have the nonzero components $e_{xx}= -e_{yy}=
{1/{\sqrt{2}}}$. This is the ``$+$'' polarization in the notation
of Ref. \cite{MTW}. Then we have that
\begin{equation}
\langle \sigma_1^2 \rangle =
        {{[(\Delta x)^2 -(\Delta y)^2)]^2}\over {16\omega V}}\,\,
{\rm Re} \bigl[\langle a^\dagger a \rangle
 + \langle a^2 \rangle e^{2i\omega(z-t)}\bigr].
\end{equation}
A squeezed vacuum state for a single mode can be defined by\cite{Caves81}
\begin{equation}
|\zeta\rangle=S(\zeta)\,|0\rangle,
\end{equation}
where $S(\zeta)$ is the squeeze operator defined by
\begin{equation}
S(\zeta) = \exp[{1\over 2}\zeta^\ast a^2
-{1\over 2}\zeta ({a^\dagger})^2],.
\end{equation}
Here
\begin{equation}
\zeta = re^{i\delta}
\end{equation}
is an arbitrary complex number. The squeeze operator has the properties that
\begin{equation}
S^{\dagger}(\zeta)\,a\,S(\zeta)=
a\,\cosh r-a^{\dagger}e^{i\delta}\sinh r,
\end{equation}
and
\begin{equation}
S^{\dagger}(\zeta)\,a^{\dagger}\,S(\zeta)=
a^{\dagger}\,\cosh r-ae^{-i\delta}\sinh r.
\end{equation}
{}From these properties, we may show that
\begin{equation}
\langle a^\dagger a \rangle = \sinh^2 r \, ,
\end{equation}
and\begin{equation}
\langle a^2 \rangle = -e^{i\delta} \sinh r \cosh r \, .
\end{equation}
Hence in this example
\begin{equation}
\langle \sigma_1^2 \rangle =
     {{\left[(\Delta x)^2 -(\Delta y)^2\right]^2}\over {16\omega V}}
 \sinh r \Bigl\{\sinh r -
\cosh r \cos\left[2\omega(z-t) +\delta\right]\Bigr\}.
\end{equation}
Here $\langle \sigma_1^2 \rangle$ will be positive in some regions and
negative in others.

      Of particular interest to us will be the case of an isotropic
bath of gravitons. Here rotational symmetry and the tracelessness
condition imply that
\begin{equation}
\langle h_{ij}h_{kl} \rangle = A\bigl(\delta_{ij}\delta_{kl} -
{3\over 2}\delta_{ik}\delta_{jl} -{3\over 2}\delta_{il}\delta_{jk} \bigr),
\end{equation}
where $A = -{1\over {15}}\langle h_{ij}h^{ij} \rangle$. In this case,
we have that
\begin{equation}
\langle \sigma_1^2 \rangle = h^2 r^4,   \label{eq:r4}
\end{equation}
where $r=|\Delta {\bf x}|$ is the magnitude of the spatial separation,
and
\begin{equation}
h^2 = -{1\over 2}A = {1\over {30}}\langle h_{ij}h^{ij} \rangle
\end{equation}
is a measure of the mean squared metric fluctuations.

    In some cases, the gravitons may be regarded as being in a thermal
state. Although a thermal state is a mixed state rather than a pure
quantum state, quantum particle creation processes often give rise to
a thermal spectrum of particles. In the case of gravitons
created by the Hawking effect, this correspondence is exact. In the case
of cosmological particle production, it is possible to obtain an
approximately thermal spectrum in some cases\cite{Parker}. We may find
$h^2$ for a thermal bath of gravitons by noting that here, due to the
two polarization states for gravitons,
$\langle h_{ij}h^{ij} \rangle = 2\langle \varphi^2 \rangle$,
where $\varphi$ is a massless scalar field. In a thermal state at
temperature $T$, it is well known that
$\langle \varphi^2 \rangle = {{T^2} \over {12}}$.
Thus, for a thermal bath of gravitons at temperature $T$,
\begin{equation}
\langle \sigma_1^2 \rangle = {1\over {180}}T^2 r^4 \,. \label{eq:thermal}
\end{equation}
Note that in this case, $\langle \sigma_1^2 \rangle >0$, whereas more generally
it may have either sign. Recall that the forms of the averaged Green's
functions obtained in Section \ref{sec:aveGF} depend upon the sign of
$\langle \sigma_1^2 \rangle$, and it is only for the case
$\langle \sigma_1^2 \rangle >0$ that expressions were found for
$\Bigl\langle G_{ret}\Bigr\rangle$ and $\Bigl\langle G_{F}\Bigr\rangle$.

\subsection{Gravitons in an Expanding Universe}
\label{sec:Grav}

         For the most part in this paper, we are concerned with gravitons
and lightcone fluctuations on a background of flat spacetime. However,
relict gravitons from the early universe are one of the more likely sources
of metric fluctuations. Thus we need to discuss gravitons on a cosmological
background, which we will take to be a spatially flat Robertson-Walker
universe. The metric can be written as
\begin{equation}
ds^2 = dt^2 -a^2(t) d{\bf x}^2 \, ,
\end{equation}
where $a(t)$ is the scale factor. Linearized perturbations of this metric
were investigated by Lifshitz \cite{Lifshitz}, who showed that it is still
possible to impose the transverse-tracefree gauge conditions,
Eq. (\ref{eq:TT}). The non-zero components of the perturbation satisfy
\begin{equation}
a^{-3} {\partial \over {\partial t}}
\Bigl(a^{3} {{\partial h^i_j} \over {\partial t}} \Bigl) -
a^{-2} \nabla^2 h^i_j =0 \, . \label{eq:perteq}
\end{equation}
However, this is just the equation satisfied by a minimally coupled
scalar field in this background,
\begin{equation}
\Box \varphi = 0 \,.
\end{equation}
Thus the graviton field may be treated as a pair (one for each polarization)
of massless, minimally coupled scalar fields. The quantization of
cosmological metric perturbations in this framework was discussed in
Ref. \cite{FP}.

     Consider a power law expansion, for which
\begin{equation}
a(t) = c t^\alpha \, .
\end{equation}
In this case, the solutions of Eq. (\ref{eq:perteq}) are of the form
$\psi_k \, e^{i{\bf k \cdot x}}$, where
\begin{equation}
\psi_k = \eta^{1\over {2b}} \bigl[c_1 H^{(1)}_\nu (k\eta) +
           c_2 H^{(2)}_\nu (k\eta) \bigr] \, .
\end{equation}
Here $b= (\alpha -1)(3\alpha -1)^{-1}$ and $\nu = (2|b|)^{-1}$. Furthermore,
$c_1$ and $c_2$ are arbitrary constants, and $\eta$ is the conformal time given
by
\begin{equation}
\eta = \int a^{-1} dt = [c(1-\alpha)]^{-1} t^{1-\alpha} \, .
\end{equation}
We are interested in the late time behavior of these solutions, which will
indicate how quantities such as $\langle \sigma_1^2 \rangle$ or $h^2$
scale with the expansion of the universe. As $t \rightarrow \infty$,
$\eta \rightarrow \infty$ if $\alpha < 1$, and $\eta \rightarrow 0$
if $\alpha > 1$. In the former case, we use the large argument limit
of the Hankel functions:
\begin{equation}
|H^{(1)}_\nu (k\eta)| \sim |H^{(2)}_\nu (k\eta)| \sim
            \sqrt{{2 \over {\pi|k\eta|}}} \,,
\end{equation}
as $|\eta| \rightarrow \infty$ for fixed $k$. In the latter case, we
use the small argument limit:
\begin{equation}
|H^{(1)}_\nu (k\eta)| \sim |H^{(2)}_\nu (k\eta)| \sim
  {{\Gamma(\nu)}\over \pi} ({1\over 2}|k\eta|)^{-\nu} \,,
      \quad |k\eta| \rightarrow 0\,.
\end{equation}
{}From these forms, we find that $|\psi_k| \sim a^{-1}$ if $\alpha <1$,
and $|\psi_k| \sim {\rm const.}$ if $\alpha >1$. Thus as
$t \rightarrow \infty$,
\begin{eqnarray}
h^2 \sim {1 \over {a^2}}\, , \qquad \alpha <1 \, , \label{eq:h1} \\
h^2 \rightarrow {\rm constant} \, , \qquad \alpha >1 \, . \label{eq:h2}
\end{eqnarray}

         Now let us make some estimates of the magnitude of $h^2$ due
to a background of relict cosmological gravitons. The creation of gravitons
in an expanding universe is a topic upon which there is a vast
literature \cite{gravrefs}. Let us consider a model
in which gravitons are created at the end of an inflationary epoch. This
type of model was discussed in Ref. \cite{F87}, where it was argued that
the typical energy density of gravitons present just after inflation will
be of the order of the energy density associated with the Gibbons-Hawking
temperature of the deSitter phase. Let $\rho_V$ be the vacuum energy
density during inflation and $\rho_P$ be the Planck density. Then the
energy density of the created gravitons at the end of inflation will be
of the order of
\begin{equation}
\rho_i \approx {{\rho_V^2}\over {\rho_P}}.
\end{equation}
This energy density will subsequently be redshifted by the expansion of
the universe to an energy density at the present time of the order of
\begin{equation}
\rho \approx \rho_i \Bigl({{3K}\over {T_R}}\Bigr)^4
     \approx {{\rho_V^2}\over {\rho_P}}\Bigl({{3K}\over {T_R}}\Bigr)^4 \, ,
\end{equation}
where $T_R$ is the temperature of reheating after inflation. Here we
are assuming that the subsequent expansion rate of the universe corresponds
to $\alpha <1$, so that the gravitons redshift as ordinary massless
particles. The typical wavelength of the gravitons at the time of creation
is
\begin{equation}
\lambda_i \approx (\rho_i)^{1\over 4}\, ,
\end{equation}
and will be redshifted at the present time to a wavelength of the order
of
\begin{equation}
\lambda \approx \lambda_i \Bigl({{3K}\over {T_R}}\Bigr)\,.
\end{equation}
The corresponding mean squared metric fluctuation will be of the order
of
\begin{equation}
h^2 \approx \rho \lambda^2 \, .
\end{equation}
If, for example, inflation were to occur at an energy scale of
$10^{15} {\rm Gev}$, and the reheating occurs to the same energy scale,
this model would predict a present-day mean graviton wavelength of the
order of $\lambda \approx 10^4 {\rm cm}$ and $h \approx 10^{-36}$. For
most purposes, the effects of these gravitons will be completely negligible.
For example, the lightcone fluctuations will produce a spread in arrival
times of pulses, from Eq. (\ref{eq:delt}), of the order of $\Delta t \approx
10^{-36} D$, where $D \leq 10^4 {\rm cm}$. This is a time spread of no more
than one Planck time and is hence unobservably small.
     The best hope for observing the effects of the lightcone fluctuations
seems to be through their indirect influence upon virtual processes, which
will be the topic of the next section.

\section{One-Loop Processes: The Electron Self-Energy}
\label{sec:oneloop}

     In this section, we wish to explore the extent to which quantum
metric fluctuations can act as a regulator of the ultraviolet divergences
of quantum field theory. These divergences typically appear in one-loop
processes, which represent the lowest order quantum corrections to the
classical theory. We will focus our attention upon the one-loop
electron self-energy. The self-energy function,
$\Sigma(p)$, is formally given by
the divergent momentum space integral:
\begin{equation}
\Sigma(p) = ie^2 \int {{d^4k}\over {(2\pi)^4}} D^{\mu\nu}_F (k) \gamma_\mu
             S_F(p-k) \gamma_\nu.
\end{equation}
Here $D^{\mu\nu}_F (k)$ and $S_F(p-k)$ are the momentum space photon and
electron propagators, respectively, and the $\gamma_\mu$ are Dirac matrices.
This integral is logarithmically divergent for large $k$. In the conventional
approach to field theory, this divergence is absorbed by mass renormalization.
Here we wish to investigate the effects of introducing metric fluctuations.
First, let us rewrite the expression for $\Sigma$ as a coordinate space
integral by use of the following relations between momentum space and
coordinate space propagators:
\begin{equation}
D^{\mu\nu}_F (k) = -\int d^4x\, e^{ikx}\,D^{\mu\nu}_F (x)\, ,
\end{equation}
and
\begin{equation}
S_F(k) = \int d^4x\, e^{ikx}\,S_F(x) \, .
\end{equation}
The electron propagator, $S_F(x)$, is expressible in terms of the scalar
propagator by the relation
\begin{equation}
S_F(x) = -(i\gamma^\mu \nabla_\mu + m_0) G_F(x). \label{eq:SFrep}
\end{equation}
Here $m_0$ might be interpreted as a bare mass.
If we adopt the Feynman gauge, the photon propagator becomes
\begin{equation}
D^{\mu\nu}_F (x) = -g^{\mu\nu} G_F(x).  \label{eq:DFrep}
\end{equation}
Note that the scalar propagator, $G_F(x)$, in Eq.~(\ref{eq:SFrep})
is that for a massive field, whereas Eq.~(\ref{eq:DFrep}) is
that for a massless field. However, we are interested in the behavior
near the classical lightcone, and so ignore the mass-dependence of the
former.
Recall that $\Sigma$ is a $4\times 4$ matrix. The mass shift can be expressed
as
\begin{equation}
\delta m = {1\over 4} {\rm Re} \bigl[Tr \Sigma(0)\bigr].
\end{equation}
If we combine the above relations and use the fact that $Tr(\gamma^\mu)=0$,
this may be written as
\begin{equation}
\delta m = m_0 e^2 \,{\rm Im} \int d^4x\, G^2_F(x).
\end{equation}
This relation has been obtained assuming a fixed, flat background metric.
However, we will assume that it also holds to leading order
when we introduce small metric perturbations.
Now we wish to average over metric fluctuations and write
\begin{equation}
\Delta m = \langle \delta m \rangle = m_0 e^2 \,{\rm Im} \int d^4x\,
                              \langle G^2_F(x) \rangle.  \label{eq:Delm}
\end{equation}

     Use Eqs.~(\ref{eq:GFsqav}) and (\ref{eq:r4}) to write
\begin{equation}
\int d^4x\, \langle G^2_F(x) \rangle =
{1 \over {(8\pi^2)}^2}
 \int_0^\infty d\alpha\, \alpha\, \int d^4x\, e^{-{1\over 2}i(t^2-r^2)\alpha}
e^{-{1\over 2} h^2 r^4  \alpha^2} \, .  \label{eq:int1}
\end{equation}
If we ignore any space or time dependence in $h$, then this integral may
be explicitly evaluated. This should be an excellent approximation, as
$h$ is expected to vary on a cosmological time scale, whereas the dominant
contributions to $\Delta m$ should come from scales of the order of or less
than
the electron Compton wavelength. If we deform the contour for the
$\alpha$-integration into the lower half plane, then  the $t$-integration
becomes absolutely convergent, and we can write
\begin{equation}
 \int_{-\infty}^\infty dt \, e^{-{1\over 2}i t^2 \alpha} =
\sqrt{\pi \over \alpha} e^{-{1\over 4}i \pi} \, .
\end{equation}
If we perform the $t$ and angular integrations in Eq.~(\ref{eq:int1}),
and then replace $\alpha$ by the variable $u = \alpha r^2$ , we find
\begin{equation}
\int d^4x\, \langle G^2_F(x) \rangle =
{\sqrt{\pi} \over {16 \pi^3}} e^{-{i\over 4}\pi} \int_0^\infty {{dr}\over r} \,
\int_0^\infty du \,\sqrt{u}\, e^{{i\over 2}u}\,
e^{-{1\over 2} h^2 u^2 } \, .  \label{eq:int2}
\end{equation}
The $r$-integration is logarithmically divergent at both limits. The infrared
divergence at large $r$ is an artifact of our having neglected the
electron mass in the electron propagator. The ultraviolet divergence at
small $r$ is more serious, and reflects the failure of metric
fluctuations to render quantum field theory fully finite. The basic problem
is that although the lightcone singularity has been removed, quantities
such as $\langle G_{F}^2 \rangle$ are still singular at coincident
points. Nonetheless, it is still of some interest to determine the
$h$-dependence of our expressions. The $u$-integration may be performed
explicitly \cite{GR1} to yield
\begin{equation}
\int d^4x\, \langle G^2_F(x) \rangle =
{{e^{-{1\over 4}i \pi}} \over {32 \pi^2}}\, h^{-{3\over 2}} \,
e^{-(16 h^2)^{-1}} \,D_{-{3\over 2}}  \Bigl(-{i\over{2h}}\Bigr) \,
                   \int_0^\infty {{dr}\over r} \,,     \label{eq:int3}
\end{equation}
where $D_p (z)$ is the parabolic cylinder function.

   If $h \ll 1$, we may use the large argument expansion \cite{GR2}
of $D_p (z)$:
\begin{equation}
D_p (z) \sim e^{-{1\over 4}{z^2}} \, z^p \,
        \biggl[ 1 - \frac{p(p-1)}{z^2} +\cdots \biggr]\,,
          \qquad |arg(z)| < \frac{3}{4} \, ,
\end{equation}
to write
\begin{equation}
D_{-{3\over 2}} \Bigl(-{i\over{2h}}\Bigl) \sim
e^{{3\over 4}i \pi}\,\, h^{3\over 2}\,
\, e^{16 h^2}\,\, (1+ 15h^2 +\cdots)\,, \qquad h \ll 1 \,.
\end{equation}
If we now combine this result with Eqs.~(\ref{eq:Delm}) and (\ref{eq:int3}),
we finally obtain the formal expression for the mass shift to be
\begin{equation}
\Delta m = \frac{m_0 e^2}{8\pi^2}\, (1+ 15h^2 +\cdots)\,
               \int_0^\infty {{dr}\over r} \, .  \label{eq:Delm2}
\end{equation}

    This expression is divergent, and hence still needs to be carefully
regularized and renormalized. Here we will simply observe that the dependence
of $\Delta m$ upon $h$ seems to be rather weak. If one were to absorb the
divergent integral into a redefinition of $m_0$, then the self-energy
would seem to be time-dependent if $h$ decreases as the universe expands.
However, this time-dependence would be extremely small at the present time.
Even if one were to identify the renormalized one-loop self energy with
the observed mass of the electron (There could be a piece of
non-electromagnetic origin.), one would have a time-dependent electron
mass with ${\dot m}/m = 30 h {\dot h}$. If $|{\dot h}/h| \approx
10^{-10}/{\rm yr}$, and $h$ is of the order of the estimate given in the
last paragraph of Sec. \ref{sec:Grav}, then  $|{\dot m}/m| \approx
10^{-80}/{\rm yr}$. This is well within the observational limits on the
time-variation of the electron mass, which are of the order of \cite{SV}
\begin{equation}
\biggl|{{\dot m} \over m}\biggr| \leq 10^{-13}/{\rm yr}.
\end{equation}

\section{Summary and Discussion}
\label{sec:summary}

       We have seen that the introduction of metric fluctuations, such as
those due to gravitons in a squeezed vacuum state, can modify the behavior of
Green's functions near the lightcone. For distinct but lightlike separated
points, the usual singularity is removed. However, the singularity for
coincident points remains. The smearing of the lightcone leads to the
possibility of ``faster-than-light light'', in the sense that some photons
will traverse the interval between a source and a detector in less than the
classical propagation time.

        The smearing of the lightcone is expected to modify virtual processes.
This was explored through the calculation of the one-loop electron self-energy
in the presence of metric fluctuations. The results were somewhat ambiguous,
due to the presence of the remaining ultraviolet divergences. They can,
however, be interpreted as supporting a very small time-dependent contribution
to the mass of the electron in an expanding universe. Of course, the dominant
source of metric fluctuations need not be relict gravitons. Any
stochastic bath of gravitons will also contribute to $h$. It is possible that
the majority of gravitons at the present time are those due to local
sources (thermal processes, etc) rather than those of cosmological origin.
It is also possible that passive metric fluctuations due to quantum
fluctuations of the energy-momentum tensor of matter produce the dominant
effect in smearing the lightcone. It would be of particular
interest to find a one-loop process which is rendered finite by the effects
of the metric fluctuations. Such a process would presumably lead to observable
quantities whose values depend upon the graviton background. Thus theories
in which gravitons regulate ultraviolet divergences can have the property that
local observable quantities may be determined by the large scale structure or
history of the universe.

\vspace{0.5cm}

{\bf Acknowledgement:} This work was supported in part by the National
Science Foundation under Grant PHY-9208805.


\begin{thebibliography}{--}
\bibitem{Boulware} The author would like to thank David Boulware for
emphasizing this point to him.
\end{thebibliography}
\end{document}